\documentclass[11pt]{article}

\usepackage{hyperref}
\pdfoutput=1

\begin{document}

\title{Mixing by Swimming Algae}

\author{Jeffrey S. Guasto$^{\ast}$, Kyriacos C. Leptos$^{\dagger}$, J.P. Gollub$^{\ast,\dagger}$,\\ Adriana I. Pesci$^{\dagger}$ and Raymond E. Goldstein$^{\dagger}$\vspace{11pt}\\
\small{$^{\ast}$Department of Physics, Haverford College, Haverford, PA 19041}\\
\small{$^{\dagger}$Department of Applied Mathematics and Theoretical Physics,}\\ 
\small{University of Cambridge, Cambridge CB3 0WA, UK}}

\maketitle

\begin{abstract}

In this fluid dynamics video (\href{http://ecommons.library.cornell.edu/bitstream/1813/13741/2/MPEG-1.mp4}{low quality}\footnote{http://ecommons.library.cornell.edu/bitstream/1813/13741/2/MPEG-1.mp4}, \href{http://ecommons.library.cornell.edu/bitstream/1813/13741/3/MPEG-2.mp4}{high quality}\footnote{http://ecommons.library.cornell.edu/bitstream/1813/13741/3/MPEG-2.mp4}), we demonstrate the microscale mixing enhancement of passive tracer particles in suspensions of swimming microalgae, \textit{Chlamydomonas reinhardtii}.\footnote{K.C. Leptos, \textit{et al.}, submitted to Physical Review Letters (2009).} These biflagellated, single-celled eukaryotes (10 $\mu$m diameter) swim with a ``breaststroke'' pulling motion of their flagella at speeds of about 100 $\mu$m/s and exhibit heterogeneous trajectory shapes. Fluorescent tracer particles (2 $\mu$m diameter) allowed us to quantify the enhanced mixing caused by the swimmers, which is relevant to suspension feeding and biogenic mixing. Without swimmers present, tracer particles diffuse slowly due solely to Brownian motion. As the swimmer concentration is increased, the probability density functions (PDFs) of tracer displacements develop strong exponential tails, and the Gaussian core broadens. High-speed imaging (500 Hz) of tracer-swimmer interactions demonstrates the importance of flagellar beating in creating oscillatory flows that exceed Brownian motion out to about 5 cell radii from the swimmers. Finally, we also show evidence of possible cooperative motion and synchronization between swimming algal cells.\footnote{Supported by NSF DMR-0803153, Leverhulme Trust, BBSRC, and Schlumberger Chair Fund.} 

\end{abstract}

\end{document}